\documentstyle[pre,multicol,aps]{revtex}
\begin{document}

\setlength{\textwidth}{180mm}
\setlength{\textheight}{240mm}
\setlength{\parskip}{2mm}

\input{epsf.tex}

\title{Two-color nonlinear localized photonic modes}

\author{Andrey A. Sukhorukov, Yuri S. Kivshar, and Ole Bang}

\address{Australian Photonics Cooperative Research Centre, 
Research School of Physical Sciences and  Engineering, \\
Optical Sciences Centre, Australian National University, 
Canberra, Australian Capital Territory 0200, Australia}

\maketitle

\begin{abstract}
We analyze second-harmonic generation (SHG) at a thin effectively 
quadratic nonlinear interface between two linear optical media. 
We predict multistability of SHG for both plane and localized waves, 
and also describe {\em two-color localized photonic modes} composed 
of a fundamental wave and its second harmonic coupled together by 
parametric interaction at the interface.
\end{abstract}


\vspace*{-1.0 cm}

\begin{multicols}{2}

\narrowtext

Cascaded nonlinearities of noncentrosymmetric optical materials have 
become an active topic of research over the last years due to their 
potential applications in all-optical switching devices 
\cite{stegeman}. 
Parametric interaction is known also to support solitary waves, 
in particular {\em spatial quadratic solitons}, which are two-frequency 
self-trapped beams consisting of a fundamental wave parametrically 
coupled to its second-harmonic \cite{sukhorukov}. 
Usually, solitary waves are considered for homogeneous media where 
spatial localization is induced by self-focusing and self-trapping 
effects. 
However, localized modes can exist even in a linear medium at 
defects or interfaces, and they are known as linear defect or interface
modes. 
The properties of {\em nonlinear defect modes} are usually analyzed 
for nonresonant Kerr-type nonlinearities \cite{braun}.  
Here we consider a qualitatively different situation and introduce 
another type of nonlinear defect mode: 
a {\em two-frequency (or two-color) localized 
photonic mode}, where the energy is localized due to the parametric 
wave mixing induced by an interface between two linear optical media 
or a thin layer with a quadratic (or $\chi^{(2)}$) nonlinearity 
embedded in a linear bulk medium.

The physical motivation for our model is twofold. First of all we point 
out a fundamental property of inhomogeneous nonlinear optical media. 
Let us consider an interface between two semi-infinite bulk 
optical media, which are either clamped together or separated by an 
infinitely thin layer.
If the bulk medium has inversion symmetry, then its quadratic 
nonlinearity must vanish. 
However, the interface breaks the symmetry and therefore 
the interface nonlinearity should possess a {\em nonvanishing 
quadratic response}, due to a nonzero contribution from the spatial 
derivatives of the electric field \cite{agran}.

Second, there exists a strong experimental evidence of SHG in localized 
photonic modes. 
For example, recent experimental results \cite{vilaseca} reported SHG 
in a truncated one-dimensional periodic photonic band-gap structure, 
in which a nonlinear defect layer was embedded. 
An enhancement of the parametric interaction in the vicinity of the 
defect was observed, suggesting that SHG occurs in local modes, while
being completely suppressed for other propagating modes. 
If the band gap of the periodic structure is wide, we can describe 
this SHG process by a model with a local quadratic nonlinear defect.

The main purpose of this paper is to introduce and study an 
analytically solvable model for SHG in nonlinear localized modes, 
based on the (induced or inherent) quadratic nonlinearity of an 
interface separating two (generally different) linear bulk media 
or a thin nonlinear defect layer embedded in a bulk medium.  
In particular, we describe {\em two-color localized nonlinear defect modes} 
where the energy localization occurs due to 
parametric coupling at the interface.

\begin{figure}
\setlength{\epsfxsize}{6.0cm}
\centerline{\mbox{\epsffile{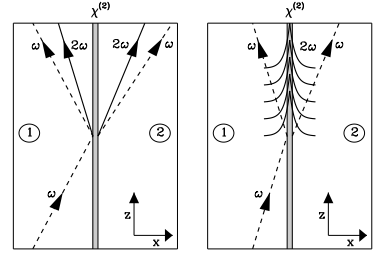}}}
\vspace{2mm}
\caption{Scattering of a plane FF wave (dashed) on a $\chi^{(2)}$-interface 
between two linear media 1 and 2.  
The generated SH (solid) can be either propagating or localized.} 
\label{fig:model}
\end{figure}

We consider a fundamental frequency (FF) wave propagating along 
the $Z$ direction in a linear slab waveguide, as shown in 
Fig.~\ref{fig:model}.
We assume that an interface (or defect layer) at $x$=0 possesses a quadratic 
nonlinear response, so that the FF wave can parametrically couple to
its second harmonic (SH) at the interface \cite{agran}. 
The coupled-mode equations for the complex envelope functions 
$E_j(x,Z)$ ($j$=1,2) can then be written in the form
\begin{equation}
\label{eq:chi2N} 
\begin{array}{l} 
  {\displaystyle  i \frac{\partial E_1}{\partial Z} + 
       D_1 \frac{\partial^2 E_1}{\partial x^2} 
         + n_1(x) E_1 + \Gamma_1(x) E_1^* E_2 = 0, } 
                   \\*[9pt]
  {\displaystyle i \frac{\partial E_2}{\partial Z} + 
       D_2 \frac{\partial^2 E_2}{\partial x^2} 
         + n_2(x) E_2 + \Gamma_2(x) E_1^2 = 0,}
\end{array}
\end{equation}
where $D_j$ are diffraction coefficients ($D_j>0$). 
For the geometry shown in Fig.~\ref{fig:model} and the approximation 
of an infinitely thin interface layer (valid when the width of the
layer is much smaller than the FF wavelength), we take 
$n_j(x) = n_{0j}(x) + \kappa_j \delta(x)$ and 
$\Gamma_j(x) = \gamma_j \delta(x)$, 
where $\gamma_j$ are the nonlinearity coefficients, 
$\kappa_j$ account for the phase velocity differences in the layer
and bulk materials, and 
$n_{0j}(x) = n_{0j}^+$, for $x>0$, and 
$n_{0j}(x) = n_{0j}^-$, for $x<0$. 

In order to reduce the number of parameters we normalize 
Eqs.~(\ref{eq:chi2N}) as follows: 
$E_1(Z)=u(z)/\sqrt{\gamma_1\gamma_2}$, $E_2(Z)=v(z)/\gamma_1$, 
$\sigma=D_2/D_1$, $\alpha_1=\kappa_1/D_1$, and $\alpha_2=\kappa_2/D_1$,
where $z=Z/D_1$ is measured in units of $D_1$. 
Then the coupled equations take the form
\begin{equation}
  \begin{array}{l} 
    {\displaystyle i \frac{\partial u}{\partial z}
 + \frac{\partial^2 u}{\partial x^2} + \nu_1 \left( x \right) u 
       + \delta \left( x \right)
         \left( \alpha_1 u + u^* v \right) = 0 , } 
                     \\*[9pt]
    {\displaystyle i \frac{\partial v}{\partial z}
 + \sigma \frac{\partial^2 v}{\partial x^2}
+ \nu_2 \left( x \right) v 
       + \delta \left( x \right)
         \left( \alpha_2 v + u^2 \right) = 0 ,}
  \end{array}             \label{eq:chi2}  
\end{equation}
where $\nu_j(x) = \nu_j^+ = n_{0j}^+ / D_1$, for $x>0$, and 
$\nu_j(x) = \nu_j^- = n_{0j}^- / D_1$, for $x<0$. 
If the mismatch $2 n_{01}^{\pm} - n_{02}^{\pm}$ is small then 
$\sigma$=1/2 is a good approximation, which we use below in the 
numerics. 
The system (\ref{eq:chi2}) conserves the Hamiltonian
\begin{eqnarray*}
 H & = & \int_{-\infty}^{+\infty}{ \left\{ 
     {\left| \frac{\partial u}{\partial x} \right|}^2 + 
     \frac{\sigma}{2} {\left| \frac{\partial v}{\partial x} \right|}^2
     - \nu_1(x) {\left| u \right|}^2
     - \frac{\nu_2(x)}{2} {\left| v \right|}^2 \right.} \\
& &  {\left. -  \delta(x){\left[  \alpha_1 {\left| u \right|}^2
            + \frac{\alpha_2}{2} {\left| v \right|}^2
            + \textrm{Re} {\left( u^2 v^* \right)} \right]}
    \right\} dx} 
,
\end{eqnarray*}
and the total 
power $P=\int_{-\infty}^{+\infty} \left( |u|^2+|v|^2 \right) dx$ for 
spatially localized or periodic solutions.

{\em Scattering problem}. To analyze the scattering process we use
that Eqs.~(\ref{eq:chi2}) are linear for $x\ne0$, and write the total 
field as a superposition of plane waves,
\begin{eqnarray}
   u(x,z)  & = & \left\{ \begin{array}{lr}
           a_1 e^{-i \lambda_1 z + i {q}_1^- x} 
         + b_1 e^{-i \lambda_1 z - i {q}_1^- x};   &  x < 0 \\
           c_1 e^{-i \lambda_1 z + i {q}_1^+ x};   &  x > 0,
       \end{array} \right.  \nonumber  \\
  v(x,z) & = & \left\{ \begin{array}{lr}
         b_2 e^{-i \lambda_2 z - i {q}_2^- x};   &  x < 0 \\
         c_2 e^{-i \lambda_2 z + i {q}_2^+ x};   &  x > 0,
       \end{array} \right. \nonumber 
\end{eqnarray}
where $a_1$, $b_1$, and $c_1$ are the amplitudes of the incident, 
reflected, and  transmitted FF waves, respectively. 
Correspondingly, $b_2$ and $c_2$ are the amplitudes of the generated 
SH waves on both sides of the interface.  
The dispersion relations are then given by
\begin{equation}
\label{eq:dispers}
  \lambda_1 = {\left({q}_1^{\pm}\right)}^2 - \nu_1^{\pm}, \;\;
  \lambda_2 = \sigma {\left({q}_2^{\pm}\right)}^2 - \nu_2^{\pm},
\end{equation}
and the continuity condition at $x$=0 yields the relations
$a_1+b_1=c_1$ and $b_2=c_2$. 
Next, integrating Eqs.~(\ref{eq:chi2}) over an infinitely small segment 
around the interface, we obtain the relations between the field derivatives 
on opposite sides of the layer,
\begin{equation}
  {\displaystyle 
    \begin{array}{rll} 
    - {\left[ \partial u / \partial x \right]}_{-0}^{+0} & = &
             {\left( u^*v + \alpha_1 u \right)}|_{x=0}, \\*[9pt]
    - \sigma {\left[ \partial v / \partial x \right]}_{-0}^{+0} & = &
            {\left( u^2 + \alpha_2 v \right)}|_{x=0}.
 \end{array} }
\end{equation}
This gives the phase-matching condition $2 \lambda_1 = \lambda_2$, which is
a general requirement for stationary propagation of FF and SH without energy
exchange, and two algebraic relations for the amplitudes,
\begin{equation}
\label{eq:cj} 
 \begin{array}{l} 
 {\displaystyle  -i \left( q_1^- + q_1^+ \right) c_1 + 2 i q_1^- a_1 = 
                 c_1^* c_2 + \alpha_1 c_1 ,} \\*[9pt]
  {\displaystyle -i \sigma \left( q_2^- + q_2^+ \right) c_2 = 
                 c_1^2 + \alpha_2 c_2. }
 \end{array}
\end{equation}
All unknown parameters $q_1^+$, $q_2^\pm$, $\lambda_j$, $b_j$, and 
$c_j$ can now be expressed in terms of the amplitude $a_1$ and 
transverse wave number $q_1^-$ of the incident FF wave. 

The SH amplitude at the interface is determined from Eq.~(\ref{eq:cj}),
\begin{equation}
 \label{eq:c2}
   c_2 = - c_1^2 / 
         \left[ \alpha_2 + i \sigma \left( q_2^- + q_2^+ \right) \right].
\end{equation}
The wave numbers are found from the phase-matching condition 
$2 \lambda_1 = \lambda_2$ and the dispersion relations (\ref{eq:dispers}),
\begin{equation}
\label{eq:om2} 
 \begin{array}{l} 
  {\displaystyle q_1^+ = \sqrt{(q_1^-)^2 + \nu_1^+ - \nu_1^-},} 
                         \\*[9pt]
  {\displaystyle q_2^\pm = \sqrt{ 
        \left[ 2 (q_1^-)^2 + \nu_2^{\pm} - 2 \nu_1^- \right] / \sigma }. }
 \end{array}
\end{equation}
These values can be either real or imaginary, corresponding
to plane waves and waves that are spatially localized at the 
interface, respectively. 
Note that the sign of the wave numbers is fixed according to 
the predefined geometry of the problem (see Fig.~\ref{fig:model}).

\begin{figure}
\setlength{\epsfxsize}{8.0cm}
\centerline{\mbox{\epsffile{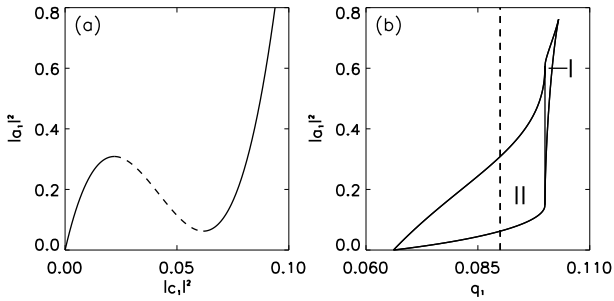}}}
\vspace{2mm}
\caption{
(a) Dependence of incident FF intensity $|a_1|^2$ on FF intensity at the
layer $|c_1|^2$, illustrating existence of multistability.
(b) Regions of multistability in $(q_1,|a_1|^2)$--space for 
$\alpha_1$=1, $\alpha_2$=0.15, and $\beta$=0.02. 
The three SH solutions are propagating (region I) or localized 
(region II). Only one solution exists outside regions I and II.
The dashed line $q_1$=0.09 corresponds to the case shown in (a).
}
\label{fig:ms_k_a1}
\end{figure}

{\em Multistability}. Substituting Eq.~(\ref{eq:c2}) into 
Eq.~(\ref{eq:cj}) we obtain the characteristic equation for the 
FF wave intensity $|c_1|^2$ at the interface, 
\begin{equation}
   |c_1^6| - 2 |c_1^4| {\rm Re}(\widetilde{\alpha}_1\widetilde{\alpha}_2) 
   + |c_1^2| \left|\widetilde{\alpha}_1\widetilde{\alpha}_2 \right|^2 
   =  4|a_1|^2|q_1^- \widetilde{\alpha}_2|^2, 
   \label{eq:c1}
\end{equation}
where 
$\widetilde{\alpha}_1 = \alpha_1 + i \left( q_1^- + q_1^+ \right)$ and
$\widetilde{\alpha}_2 = \alpha_2 + i \sigma \left(q_2^- + q_2^+ \right)$. 
Equation~(\ref{eq:c1}) is cubic in 
$|c_1|^2$, and thus {\em three different roots} may exist 
for a given input intensity $|a_1|^2$, corresponding to three 
different values of the amplitudes at the interface,
as shown in Fig.~\ref{fig:ms_k_a1}(a). This describes 
a {\em multistable} SHG process. 

To study the stability of these solutions we investigate the corresponding
linearized problem, similar to the analysis of a different problem in
Ref.~\cite{malomed}. It is convenient to chose the perturbation functions
with profiles which remain self-similar upon propagation in the
inhomogeneous medium. Then it can be demonstrated that the growth rate 
for nonoscillatory instability modes vanishes at the turning points 
$\partial |a_1|^2 / \partial |c_1|^2 = 0$.
We have verified numerically that the growth rate is positive on the branch 
with negative slope [dashed line, Fig.~\ref{fig:ms_k_a1}(a)], 
meaning that the corresponding solutions are unstable.
However, a rigorous stability analysis of all nonlinear modes is beyond 
the scope of this paper.

Let us consider the simplest possible example, in order to illustrate the
characteristic physical properties of the system.
We choose the case when the linear media 1 and 2 on each side of the interface 
(see Fig.~\ref{fig:model}) are identical, i.e., $\nu_j^\pm=\nu_j$.
It then immediately follows from Eqs.~(\ref{eq:om2}) that 
$q_j^{\pm} = q_j$. 
This means that the SH can exist in two different states, 
propagating or localized, whereas the FF waves are always propagating.
The localized SH state can only be observed if
$\beta > 0$, and then only for FF wave numbers less than a critical value:
$q_1 < \sqrt{ \beta / 2}$, where $\beta \equiv 2 \nu_1 - \nu_2$. Note that
localization does not depend on the wave amplitudes.

For fixed material parameters $\alpha_j$ and $\beta$ the multistable 
SHG regime can only be observed in certain regions of the input 
parameters $q_1$ and $|a_1|^2$, as illustrated in 
Fig.~\ref{fig:ms_k_a1}(b).
Importantly, multistability can be found for both propagating (region I) 
and localized (region II) SH waves.
For other values of the material parameters the multistable scattering
might be observed for a single type of the SH waves, or only a
single-state field configuration may be possible.

\vspace{0.1 cm}

\begin{figure}
\setlength{\epsfxsize}{8.0cm}
\centerline{\mbox{\epsffile{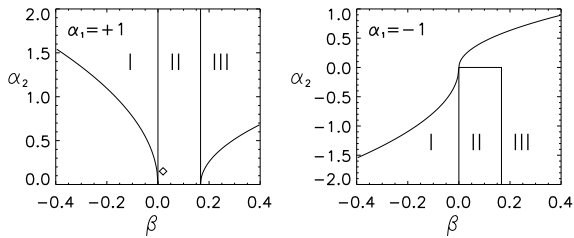}}}
\vspace{2mm}
\caption{Regions I-III in ($\beta,\alpha_2$)--space for $\alpha_1=\pm$1
where multistability can be observed. 
In region I all three SH states are propagating, in II they 
can be either propagating or localized, and in III they are all
localized. The marked point corresponds to the case presented in
Fig.~\ref{fig:ms_k_a1}.}
\label{fig:ms_al2_beta}
\end{figure}

In Fig.~\ref{fig:ms_al2_beta} we summarize the different SHG regimes
in terms of the material parameters.
The diagrams are presented for $\alpha_1 = \pm 1$, but they are invariant 
to the scaling
$\beta \rightarrow \beta / \alpha_1^2$, 
$\alpha_2 \rightarrow \alpha_2/ |\alpha_1| $, meaning that they can 
characterize the multistability regions for any $\alpha_1$.

\vspace{0.1 cm}

\begin{figure}
\setlength{\epsfxsize}{8.0cm}
\centerline{\mbox{\epsffile{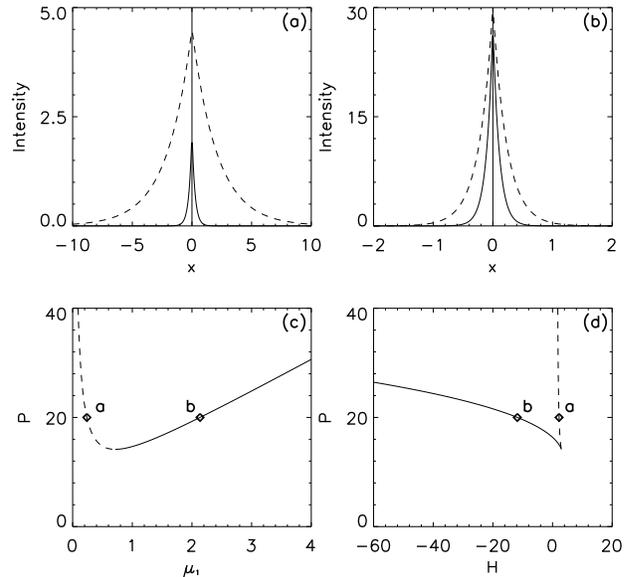}}}
\vspace{2mm}
\caption{(a,b) Intensity profiles of two different localized states 
with the same power $P$=20 (FF -- dashed, SH -- solid). 
(c) Power $P(\mu_1)$ with two branches.
(d) Hamiltonian $H$ vs. power $P$. The unstable branch is depicted with a 
dashed line. Points a and b mark the modes shown in (a) and (b), respectively.
For all the plots $\alpha_1 = \alpha_2=-$1, $\beta = 2$.}
\label{fig:loc_e_beta}
\end{figure}

{\em Nonlinear localized modes}. 
As mentioned above, in the scattering of a plane FF wave 
the transmitted FF and generated SH waves 
can be either propagating or localized. 
However, the situation when all the FF and SH waves are localized is 
also possible.
These {\em two-frequency nonlinear localized modes} are of
significant physical interest.
To find stationary solutions for these modes we assume the following 
conditions in the general scattering problem:  
(i) no incident plane wave, i.e.,~$a_1$=0, and 
(ii) all transverse wave numbers are imaginary, 
$q_j^ \pm = i \mu_j^{\pm}$, where $\mu_j^{\pm}$ are real and positive (as
the wave amplitudes should vanish at infinity). 
Then the amplitudes at the interface are 
$|c_1|^2 = \left( \mu_1^- + \mu_1^+ - \alpha_1 \right)
           \left[ \sigma \left( \mu_2^- + \mu_2^+ \right) - \alpha_2 \right]$
and 
$c_2 = \mu_1^- + \mu_1^+ - \alpha_1$. Note that here only one
wave number is arbitrary, all others are determined by Eq.~(\ref{eq:om2}).
Such localized states can exist for any combination of material parameters.
Some examples are presented 
in Figs.~\ref{fig:loc_e_beta}(a) and \ref{fig:loc_e_beta}(b).

For the symmetric case, where $\nu_j^{\pm} = \nu_j$ and $\mu_j^{\pm} = \mu_j$, 
the total power $P = |c_1|^2 / \mu_1 + |c_2|^2 / \mu_2$ can be written as a
function of $\mu_1$ only. 
The dependence $P(\mu_1)$ has always, for any values of the material 
parameters, a branch with positive slope.
Under certain conditions a second branch with negative slope may appear. 
This branch corresponds to smaller values of $\mu_1$ and larger values of
the Hamiltonian $H$, as shown 
in Figs.~\ref{fig:loc_e_beta}(c) and \ref{fig:loc_e_beta}(d) 
with dashed lines. Thus we may
conclude that for two bistable states the one with lower $\mu_1$ 
(i.e., higher $H$ and negative slope ${\partial P / \partial \mu_1} < 0$) is
unstable. For other values of the material parameters the power ranges
corresponding to branches with positive and negative slope may not fully
overlap, or there can even be a gap.

\vspace{0.1 cm}

Let us consider the generation of a stable two-color localized mode 
by launching a localized FF wave at the interface.
As the first step in the analysis of the dynamical problem 
we consider a simplified case,
assuming that both the amplitude and phase velocity of the FF 
pump wave at the interface are constant. 
This is a reasonable approximation, if the initial FF 
wave is close to a stationary mode localized due to a linear phase 
detuning at the layer (characterized by $\alpha_1$), and
the generated SH wave remains small (essentially an undepleted pump
approximation). For such a case, the original 
system  (\ref{eq:chi2}) can be reduced to a single equation for 
the SH wave,
\begin{equation}
\label{eq:reduction}
  i \frac{\partial v}{\partial z} 
     + \sigma \frac{\partial^2 v}{\partial x^2} 
     + \nu_2 v 
     + \delta(x) \left( \alpha_2 v +
                {\left| c_{10}^2 \right|} e^{- 2 i \lambda_1 z} \right)
     = 0 ,
\end{equation}
with the initial condition $v(0)$=0. 
In Eq.~(\ref{eq:reduction}) the initial intensity of the FF wave 
at the interface $|c_{10}^2|$ and its propagation constant $\lambda_1$
are fixed.

\begin{figure}
\setlength{\epsfxsize}{8.0cm}
\centerline{\mbox{\epsffile{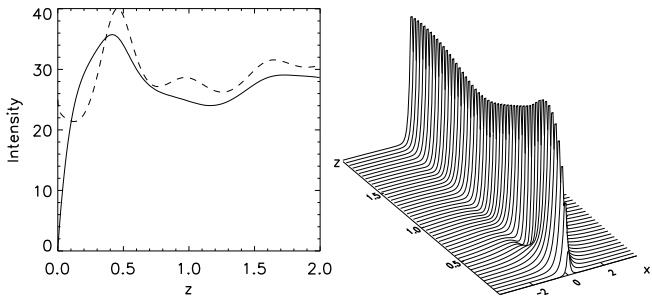}}}
\vspace{2mm}
\caption{Generation of a two-color localized mode (material parameters are
the same as in Fig.~\ref{fig:loc_e_beta}). 
Left: intensities of the FF (dashed) and SH (solid) modes at the interface. 
Right: evolution of the excited SH. 
The input power is $P \simeq 28$ and the final state is 
close to that shown in Fig.~\ref{fig:loc_e_beta}(b).}
\label{fig:loc_shg}
\end{figure}

In the case $\alpha_2$=0, an exact solution of Eq.~(\ref{eq:reduction}) 
can be presented in the form,
\[  
  v(x,z) = e^{ - 2 i \lambda_1 z}
      \int_0^z
    {\frac{i {\left| c_{10}^2 \right|}}{2 \sqrt{i \pi \sigma \zeta}} 
      e^{i \left( 2 \lambda_1 + \nu_2 \right) \zeta}
      e^{ {i x^2} / {4 \sigma \zeta} } d\zeta } .
\]
The expression under the integral describes decaying oscillations with
the increase of $z$. Thus, the amplitude of the generated SH exhibits
oscillations as the solution approaches asymptotically a stationary 
two-color localized state.
We have performed a number of numerical simulations 
(using a fully implicit finite-difference method)
with Gaussian initial
profiles of the FF wave and found that 
in the general case the formation of localized modes is  
accompanied by the same kind of transitional oscillations, as shown in 
Fig.~\ref{fig:loc_shg}. Furthermore, we also observed switching from a
perturbed unstable two-color localized mode, 
such as that shown in Fig.~\ref{fig:loc_e_beta}(a), to a stable one.
In the evolution process some energy is radiated, and the 
power corresponding to the localized mode is decreased
accordingly. Thus, a stable localized mode can only be generated, if initial
power is above the threshold $P_{\rm th} = {\rm min}_{\mu_1} P( \mu_1 )$.

\vspace*{-0.4 cm}

In conclusion, we have introduced an analytically solvable model for 
SHG in localized modes and predicted
the existence of two-color nonlinear localized photonic modes 
supported by parametric interaction at an interface. 
Some of the properties of two-color localized modes, such as
stability, generation, and switching, 
show a remarkable similarity with parametric solitons in 
homogeneous optical $\chi^{(2)}$ media \cite{sukhorukov} and their 
interaction with localized perturbations of the mismatch parameter 
\cite{prl_torner}. 
We believe that these 
results open a new class of problems in the theory of nonlinear wave 
propagation in inhomogeneous media with resonant nonlinearities, and 
they should be useful to understanding the fundamental difference between 
the effects produced by nonresonant Kerr-type nonlinearities and 
those induced by parametric wave interaction. 
For example, the analysis of localized modes due to a single 
$\chi^{(2)}$ defect is the first step towards the theory of 
nonlinear modes and gap solitons in  nonlinear photonic crystals 
\cite{berger} and the dynamics of the defect modes in such materials.

~\\
The authors are indebted to C.~Soukoulis and R.~Vilaseca for useful
discussions and comments.
The work was partially supported by the Department of Industry, 
Science, and Tourism (Australia).

\end{multicols}


\begin{references}

\bibitem{stegeman} 
   For an overview, see G. Stegeman, D.J. Hagan, and L. Torner,  
   Opt. Quantum Electron. {\bf 28}, 1691 (1996).

\bibitem{sukhorukov} 
   Yu.N. Karamzin and A.P. Sukhorukov, Zh. Eksp. Teor. 
   Fiz. {\bf 68}, 834 (1975) [Sov. Phys. JETP {\bf 41}, 414 (1976)]; 
   A.V. Buryak and Yu.S. Kivshar, Phys. Lett. A {\bf 197}, 407 (1995); 
   D.E. Pelinovsky, A.V. Buryak, and Yu.S. Kivshar, 
   Phys. Rev. Lett. {\bf 75}, 591 (1995); 
   L. Torner, In : {\em Beam Shaping and Control with Nonlinear 
   Optics},  
   F. Kajzar and R. Reinisch, Eds. (Plenum, New York, 1998), p. 229.

\bibitem{braun} 
   For a general overview, see 
   O.M. Braun and Yu.S. Kivshar, Phys. Rep. {\bf 306}, 
   1 (1998), and references therein.

\bibitem{agran} 
   V.M. Agranovich and S.A. Darmanyan, JETP Lett. {\bf 35}, 
   80 (1982) [Pis'ma Zh. Eksp. Teor. Fiz. {\bf 35}, 68 (1982)].

\bibitem{vilaseca} 
   J. Trull, R. Vilaseca, J. Martorell, and R. Corbal\'an, 
   Opt. Lett. {\bf 20}, 1746 (1995).

\bibitem{malomed}
   B.A. Malomed and M.Ya. Azbel, Phys. Rev. B
   {\bf 47}, 10402 (1993).

\bibitem{prl_torner} 
   C. Balslev Clausen and L. Torner, Phys. Rev. Lett. 
   {\bf 81}, 790 (1998); 
   C. Balslev Clausen, J.P. Torres, and L. Torner, Phys. Lett. A 
   {\bf 249}, 455 (1998).


\bibitem{berger} 
   V. Berger, Phys. Rev. Lett. {\bf 81}, 4136 (1998).

\end{references}
\end{document}